\documentclass[aps,preprintnumbers,epsfig,prl,final,twocolumn,showpacs,superscriptaddress]{revtex4}
\usepackage{amssymb}
\usepackage{amsmath}
\usepackage{amsopn}
\usepackage{graphics}
\usepackage{graphicx}
\usepackage{epsfig}

\begin{document}

\title{Electron acceleration by colliding laser beams in plasmas}

\author{A. F. Lifschitz}
\affiliation{Laboratoire d'Optique Appliqu\'{e}e, ENSTA/Ecole
Polytechnique (CNRS UMR 7639), 91761 Palaiseau cedex, France}
\affiliation{Laboratoire de Physique des Gaz et des Plasmas (CNRS UMR 8578), Universit%
\'{e} Paris XI, B\^{a}timent 210, 91405 Orsay cedex, France}
\author{ J. Faure}
\author{ C. Rechatin}
\author{ V. Malka}
\affiliation{Laboratoire d'Optique Appliqu\'{e}e, ENSTA/Ecole
Polytechnique (CNRS UMR 7639), 91761 Palaiseau cedex, France}
\author{ E. Lefebvre}
\affiliation{D\'epartement de Physique Th\'eorique et Appliqu\'ee, CEA/DAM Ile-de-France, BP 12, 91680 Bruy\`eres-le-Ch\^atel, France}

\date{\today}
\preprint{}

\begin{abstract}
All-optical controlled injection and acceleration of electrons in
a laser wakefield has recently been achieved \cite{faure2006}.
Injection was made using a second  counterpropagating laser pulse
with the same polarization as the main pulse . In this scheme, the
interference pattern formed during the collision between the
pulses pre-accelerates electrons that are then trapped in the
wakefield. Numerical simulations of this process performed with a
new Particle-in-Cell code are presented here. The results show the
nature of the injection mechanism and explain some striking
features found experimentally, such as the dependence of beam
energy with the collision position and the reduction of the charge
at high energies. Simulations reproduce well the main features of
the experimental observations.

\end{abstract}

\maketitle

The field of laser-plasma wakefield acceleration has known a fast
development over the past few years. In this approach, particles are accelerated in the
relativistic plasma wave excited by an intense laser propagating in a
underdense plasma. Electric fields in the wake can reach some hundreds
of GeV/m instead of the tens of MeV/m attainable in conventional RF
accelerators,  opening the door to a new generation of compact
particle accelerators. Up to now, a single laser beam was responsible
for the injection and subsequent acceleration of the electrons \cite{malka2002,mangles2004,geddes2004,faure2004}.  Very recently,
the external injection of electrons in the wakefield using a second
laser beam has been demonstrated \cite{faure2006}. Electron beams
obtained in this manner are quasi-monoenergetic, tuneable and stable.

The idea of using a second laser to inject the electrons was
proposed years ago \cite{umstadter1997}, and further developed in
the scheme implemented in the experiments \cite{esarey1997}.  In
this version, two counterpropagating laser pulses with the same
wavelength $\lambda_0$ and polarization are used. The first pulse
(the pump pulse), with normalized amplitude $a_0>1$, creates a
wakefield. The second pulse (the injection pulse), with normalized
amplitude $a_1<1$, collides with the pump pulse.  During the
collision, a laser beatwave pattern with phase velocity
$v_{bw}\approx 0$ is formed \cite{esarey1997}. The scale of  the
beatwave pattern is $\lambda_0/2$, therefore the ponderomotive
force of the beatwave $F_{bw} \propto 2 a_0 a_1/\lambda_0$ is very
large. This very large  ponderomotive force pre-accelerates 
electrons. A fraction of them is then trapped in the wakefield (depending
on the values of $a_0$ and $a_1$) and subsequently accelerated to
relativistic energies.

Previous numerical studies of colliding pulse injection deal with
fluid descriptions of the plasma (valid for normalized laser
amplitudes $a_0 < 1$) and test particle treatment for the injected
electrons \cite{fubiani2004}, or one-dimensional kinetic
simulations \cite{kotaki2004}. We will show that a realistic
description of the process requires to include 3D beam dynamics
issues,  including self-focusing \cite{sun1987} and
self-compression \cite{faure2005} of the pump pulse as well as a
correct description of the electric radial wake field and beam
loading effects. In previous studies  these effects were absent or
not well described due to the use of low dimensions.

In order to deal with this problem without a full
three-dimensional kinetic simulation, we have developed a new
Particle-in-Cell code with an hybrid  cylindrical-geometry scheme.
Maxwell equations are written in axisymmetric cylindrical
coordinates ($r,z$), with $z$ the direction of propagation of the
laser beams.  Particles evolve in tridimensional space.  To
calculate the sources of the Maxwell equations (the density
$n(r,z)$ and the current $j(r,z)$), tridimensional particle
information, i.e. position ($x,y,z$) and velocity ($v_x,v_y,v_z$),
is projected over the $(r,z)$ grid.

The laser pulses are not described as ordinary
electromagnetic fields but via their envelope, $a(r,z)$.
Each laser beam is taken as a
monochromatic plane wave of frequency $\omega_0$ and wavenumber
$k_0=\omega_0/c$ modulated by the envelope amplitude $a(r,z)$.
To calculate the laser electromagnetic field over a given particle, we
interpolate the  amplitudes of the pump and injection beams
 ($a_0^i$ and $a_1^i$ respectively) from the grid over the particle
position ($x,y,z$), and multiply this by the high-frequency
component. If both pulses are counterpropagating and linearly
polarized in $\hat x$, the normalized electric and magnetic laser
fields are given by
\begin{eqnarray}
E_x^{L}&\!\!\!\!= a_0^i \sin(k_0  z+ \omega_0 t+\phi_0)
\! + \! a_1^i \sin(k_0 z-\omega_0 t+\phi_1)&  \nonumber \\
B_y^{L}&\!\!\!\!= a_0^i \sin(k_0 z+ \omega_0 t+\phi_0)\! - \!a_1^i \sin(k_0 z-\omega_0 t+\phi_1),
\end{eqnarray}
with $\phi_0$ and $\phi_1$ arbitrary phases, that we set to zero.
Note that the high-frequency laser field is not calculated over
the grid but over each particle, so that the mesh size can be
larger than $1/k_0$, reducing drastically the computational cost.
On the other hand, the timestep is limited by $1/\omega_0$. When
the pulses do not overlap, the wakefield created by each laser
beam is similar to the one obtained using the ponderomotive force
$F \propto\nabla |a(r,z)|^2$, i.e. the  wakefield is axially
symmetric. Therefore, the assumption of axial symmetry constitutes
a valid approximation. However, the inspection of electrons
trajectories during the beatwave shows that the symmetry is
partially broken in the collision region. The spatial distribution
of electrons trapped in the vicinity of the laser pulse also
exhibits a degree of asymmetry \cite{mangles2006}. We will discuss
later the importance of this lack of symmetry.

\begin{figure}[t]
\begin{center}
\resizebox{7.0cm}{!}{\includegraphics{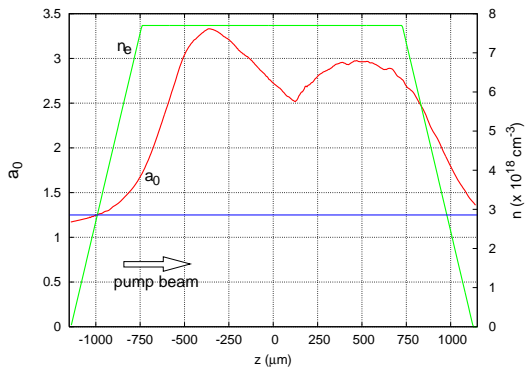}}
\end{center}
\caption{  Electronic density profile ($n$) and evolution of the normalized amplitude of the pump beam
  ($a_0$)  along the gas jet. } \label{a0_vs_z}
\end{figure}

The pump pulse envelope  is calculated using the  envelope
equation given in \cite{mora1997}. The injection pulse is slightly
self-focused (the maximum amplitude is $a_1\approx 0.6$ instead of
$a_1=0.4$). For the sake of numerical simplicity, we have
neglected self-focusing of the injection pulse and have considered
that it is the analytical solution of the envelope equation in
vacuum. This is justified because simulations show that injection
is weakly dependent on the amplitude of the injection pulse in the
range $a_1=0.4-0.7$.

Parameters of the simulations we have performed correspond to the
experiments recently performed at LOA \cite{faure2006}. The pump
laser energy is 700 mJ, $\lambda_0=2 \pi/k_0=0.8~\mu$m, pulse
duration 30 fs and spot size FWHM 16 $\mu$m, corresponding to an
intensity $I=3.4\times 10^{18} ~\mbox{W/cm}^2$, a normalized
amplitude $a_0=1.25$ and a Rayleigh length $z_R=1.5$ mm. The
injection pulse energy is 250 mJ, $\lambda_0=0.8~\mu$m, pulse
duration 30 fs and spot size FWHM 30 $\mu$m, corresponding to
$I=4\times 10^{17} ~\mbox{W/cm}^2$,
 $a_1=0.4$ and $z_R=4$ mm.
The plasma density  profile (figure \ref{a0_vs_z})  is close  to
the experimental one. The evolution of the pump beam amplitude as
the beam propagates along the plasma is shown in figure
\ref{a0_vs_z}. The beam propagates from left to right and its
focal plane is at $z=-750 ~\mu$m (the center of the jet is at
$z=0$). The focal amplitude  in vacuum ($a_0=1.25$) is indicated
by the horizontal straight line. As we can see, the pump beam
start being self-focused in the density ramp at the beginning of
the jet, reaching a maximum of $a_0 \simeq 3.4$ at $z=-300 ~\mu$m.
From this point, the beam diverges and it is again self-focused in
the second half of the jet. The density in the exit ramp of the
jet is too low to counter beam diffraction. The pulse is also
self-compressed, reaching  a duration of about 20 fs at the
exit.

Simulations made  without the injection beam indicate the
existence of a small degree of self-injection. The spectrum of
self-injected electrons is  flat between 20 and 300 MeV , with a
 total charge of $\simeq 7$ pC. Electrons are trapped  far from the
laser,  beyond the third wake period.

\begin{figure}[t]
\begin{center}
\resizebox{6.5cm}{!}{\includegraphics{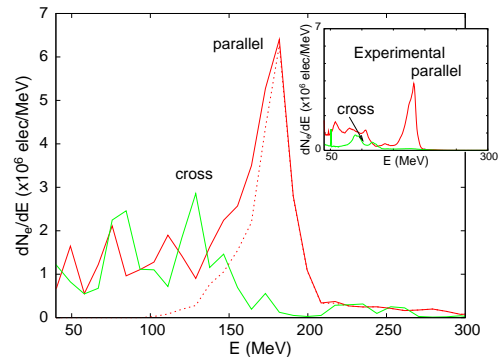}}
\end{center}
\caption{ Numerical spectra for parallel and cross polarization.  Beam collision
 takes place at $z=-650~\mu$m.  Dotted line is the spectrum of electrons trapped in
  the bubble (parallel polarization).  Experimental spectra for
 parallel and cross polarization are shown in the inset plot.
  } \label{par_per}
\end{figure}

When an injection beam with parallel polarization is included, a
fraction of the electrons present during the beatwave between the
pulses is trapped in the wakefield, its total charge ranging
between 50 pC and 200 pC. Most of the electrons are trapped in the
bubble (we use the term ``bubble'' to refer to the first bucket of
the wakefield structure, although in the present study laser
intensities are not high enough to reach the bubble regime), and
they contribute to the energy spectrum as a high energy narrow
peak  (figure \ref{par_per}). When the polarizations are crossed,
it is also possible to trap some electrons, although the high
energy peak disappears and the charge is divided by $\sim 3$. These electrons are grouped in several
bunches and they  are far from the laser (the closest bunch is in
the fourth wake period). The energy spectrum is broad, with
several peaks corresponding to electrons trapped in different
bunches (figure \ref{par_per}).
 A similar trend has been observed in the experiments, i.e. spectra with a
high energy peak for parallel polarization and lower charge
multi-peaked or wide spectra for cross polarization (inset in
figure \ref{par_per}). A good agreement between simulation and
experiment is found. However, the charge is slightly overestimated
in the simulations as well as the relative energy spread.
\begin{figure*}[t]
\begin{center}
\resizebox{14cm}{!}{\includegraphics{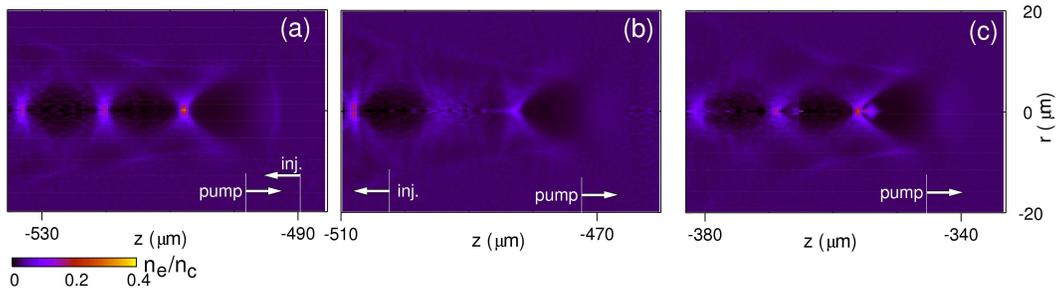}}
\end{center}
\caption{Spatial distribution of electron density for parallel
  polarization at three times: before the collision
  (a), shortly after the collision (b) and for a later time (c). Collision
  take place at $z_c=-495~\mu$m.
  } \label{dens_par}
\end{figure*}

The trapped charge for parallel polarization is small compared
with predictions made by previous models \cite{fubiani2004}. The
reason of this low charge  is that the beatwave is strong enough
to locally destroy the wakefield. Most of
 the particles that were pre-accelerated by the beatwave are not trapped simply
 because during the collision, there is no accelerating nor radial field to trap
 them. Only the electrons accelerated at the end of the
 collision reach the pulse back late enough to find a newly
 generated wakefield structure. Figure \ref{dens_par} illustrates this
 phenomenon. It shows the spatial distribution of electronic density before the collision
 (\ref{dens_par}.a), shortly after the collision (\ref{dens_par}.b) and
 for a later time (\ref{dens_par}.c). As we can see, before the
 collision a periodic wake structure exists. Once the collision
 takes place, the region of the wakefield where the beatwave occurred
 is strongly distorted. The short scale ($L \sim \lambda_0/2$) ponderomotive force associated
 with the beatwave is stronger than the long range  ($L \sim
 \lambda_p$) ponderomotive force of the
 pump laser, and no coherent movement of the electrons in the scale of
 $\lambda_p$ takes place. Therefore, the plasma wave is not excited in
 this region and no wake is formed. As the pump beam moves away from
 the collision region,  the wake is reformed and the  bunch of electrons
 trapped in the bubble becomes clearly visible (figure \ref{dens_par}.c). We have made simulations in which the longitudinal and transverse wakefield were artificially
 frozen shortly before the collision, i.e. the wakefield is unaffected by
 the beatwave. We have found that the charge injected in the bubble is multiplied by a
 factor $\sim 10$. For example,  for collision at $z_c=-650 ~\mu$m, the charge in the bubble raises from 30 pC to 300 pC.

\begin{figure}[b]
\begin{center}
\resizebox{6.5cm}{!}{\includegraphics{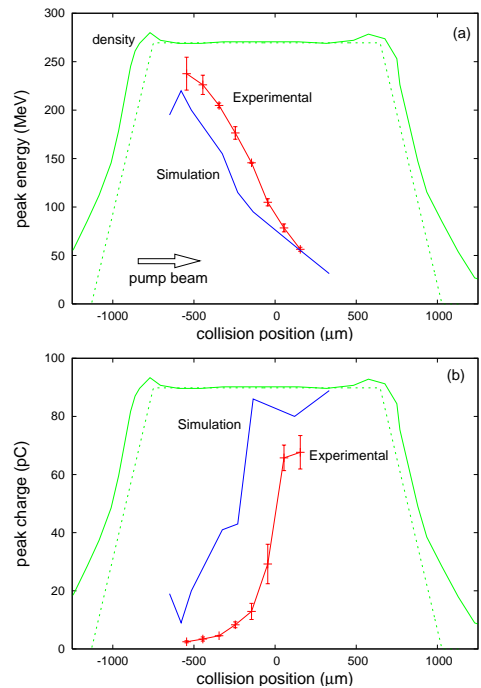}}
\end{center}
\caption{Numerical and experimental peak energy (a) and peak charge (b) vs. collision position.} \label{e_q}
\end{figure}

In the following we compare the numerical predictions for parallel
polarization with experimental results. Figure \ref{e_q}.a shows
the energy of the monoenergetic peak as a function of the
collision position, along with the experimental results. When the
spectrum presents more than one peak, the highest energy peak is
chosen. Except for the earliest collision case ($z_c=-650 ~\mu$m),
the final energy decreases with $z_c$, i.e. a shorter acceleration
distance corresponds to a lower final energy. We can see that
there is a good agreement between the simulations and the
experimental data, even though the theoretical curve is shifted to
the left by $\sim 150-250 ~\mu$m.

When the collision between the pump and the injection beams takes
place close to the beginning of the  plateau ($z_c<-650 ~\mu$m),
no extra energy gain can be obtained. In this region, the pump
pulse amplitude grows quickly (figure \ref{a0_vs_z}). As the laser
field becomes stronger, the plasma wavelength progressively
stretches due to the relativistic redshift of the plasma
frequency, i.e. the wakefield slows down \cite{lifschitz2005}.
Trapped electrons are then able to overcome the wakefield,  moving
farther from the maximum of accelerating field.

The experimental charge in the monoenergetic peak (figure
\ref{e_q}.b) decreases with the peak energy, and it exhibits a
drastic drop above 100 MeV. The same trend is found in the
simulations, although the charge is overestimated by a factor
between $\sim 1.3 - 3$. Here again, the theoretical curve is
shifted to lower $z_c$. Simulations show
  that the number of trapped electrons remains almost unchanged after
  the collision. Therefore, the reduction of the charge with the energy is not due to an
  extra loss of trapped electrons along the longer acceleration
  path.

The injected charge is firstly determined by the pump laser
amplitude ($a_0$) that grows between $z=-1000$ and $-300~\mu$m
due to self-focusing (figure \ref{a0_vs_z}). For $z_c>-300~\mu$m,
the trapped charge continues to grow even when the laser amplitude
starts to decrease. This is due to the  progressive
self-compression of the pump pulse, that drives the wake more
efficiently. Besides this, simulations show that for a given
amplitude of the longitudinal field, the self-compressed and
distorted pulse is able to trap more electrons.
 Further study is required to understand the larger trapping of
self-compressed pulses.

A second factor can play a role in the steep drop of the peak
charge above 100 MeV. At high energies ($>100$ MeV), the peak
corresponding to the electrons in the bubble is clearly distinct
from  the contribution of the other electrons. At low energies,
these contributions partially overlap, originating a larger peak.

\begin{figure}[tbph]
\begin{center}
\resizebox{6.5cm}{!}{\includegraphics{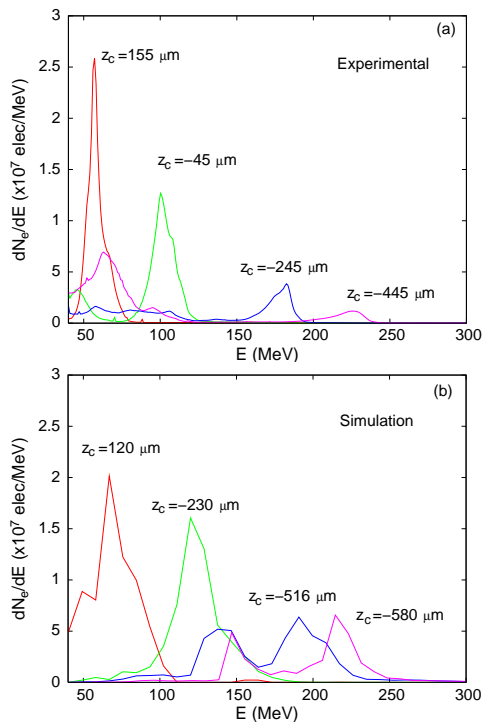}}
\end{center}
\caption{Experimental spectra (a) and simulation results (b) for selected
  collision positions.} \label{spec}
\end{figure}

Figure \ref{spec} shows experimental spectra and simulation
results for selected collision positions. Spectra includes
typically the high energy peak tuned by the collision position
plus a low energy component. In the simulations, the high energy
peak corresponds to the electrons in the bubble. For low energies
($E \le 100$ MeV), the contribution from electrons inside and
outside the bubble overlaps, resulting in a single peak . The
separation between the high energy peak and the low energy region
is much clearer in the experimental spectra than in the
theoretical curves.

The simulations we have presented underline the role played by a
number of effects in the determination of the electron beam
properties.
  The increase of laser intensity due
to self-focusing accounts for the
very high acceleration gradients inferred from experimental
data. In addition to that, self-focusing, along with
 pulse shortening due to self-compression, enhances the trapping and therefore, the charge of
the accelerated bunches. On the other hand, the wake destruction
due to the beatwave reduce drastically the accelerated charge.
According to simulations, the high energy quasi-monochromatic
peaks found experimentally correspond to a single and short
electron bunch accelerated in the bubble. Bunch length predicted
by the simulations ranges between 10 fs FWHM at 60 MeV and 7 fs
FWHM at 220 MeV.

The good agreement between simulations and experiments indicates
that the assumption of axial symmetry constitutes a good
approximation. Nevertheless, this approximation can be partly
responsible for the overestimation of the charge in the
simulations, because the projection of particle density over
($r,z$)  could reduce to some extent the degree of wake
destruction caused by the beatwave. The second source of asymmetry
is the presence of energetic particles in regions where the laser
field is significant. Due to the relatively modest charge of the
bunches in the bubble as well as the shortening of the pump pulse
due to self-compression (that reduces the laser intensity over the
bunch), the relevance of this effect should be small.

We acknowledge the support of the European
Community Research Infrastructure Activity under the FP6
Structuring the European Research Area program (CARE, contract
number RII3-CT-2003-506395).

\end{document}